\documentclass[twocolumn,showpacs,preprintnumbers,amsmath,amssymb]{revtex4}

\usepackage{graphicx}
\usepackage{psfrag}
\usepackage{dcolumn}
\usepackage{bm}

\newcommand{\RR}{\mathbb{R}}
\newcommand{\beq}{\begin{equation}}
\newcommand{\eeq}{\end{equation}}

\begin{document}

\preprint{APS/123-QED}

\title{Transition from stable orbit to chaotic dynamics in hybrid systems of Filippov type with digital sampling}

\author{Paul Glendinning}
\author{Piotr Kowalczyk}%
 \email{piotr.kowalczyk@manchester.ac.uk}
\affiliation{
Centre for Interdisciplinary Computational and Dynamical Analysis (CICADA) and School of Mathematics, \\ University of Manchester,
Oxford Road, Manchester M13 9PL, U.K.
}%

\date{\today}

\begin{abstract}
We demonstrate on a representative example of a planar hybrid system with digital sampling a sudden transition from a stable limit cycle to the onset of chaotic dynamics.
We show that the scaling law in the size of the attractor is proportional to the digital sampling time $\tau$ for sufficiently small values of $\tau.$ Numerical and analytical results are given. The scaling law changes to a nonlinear law for large values of the sampling time $\tau.$ This phenomenon is explained by the change in the boundedness of the attractor.
\end{abstract}

\maketitle

The control, design and analysis of many systems relevant to real world applications involves understanding
the interaction between continuous and discrete dynamics. For example, the automated control
of a car moving on a road is implemented by
digital computer but the motion of a car is continuous in time \cite{GuUtAc:94}. Hence the
design of the control of such a system needs to take into account the effects of the interaction between
the continuous and discrete dynamics.  Other examples include the control of the motion of digitally
controlled machines, for instance in robotics \cite{HaSt:96,EnSt:98,KoStHo:00}. These types of control systems are often referred
to in the control literature as hybrid control systems \cite{Li:06}. In \cite{HaSt:96,EnSt:98} it has been shown that the
digitization of the spatial structure by the controller induces micro-chaotic transient dynamics. Effects of digitization on the stability of the solutions have been considered in \cite{LeHa:02,BrKoFe:06}, and in \cite{XiCh:07} the existence of different types of attractors in a simple model of a delta-modulated control system has been shown.
Here we consider another aspect of digitization in hybrid systems. We assume that the input to the
controller is delivered at discrete times, separated by a constant $\tau>0$, and show that for
arbitrarily small $\tau$ the system can exhibit chaotic dynamics.
Moreover, there are scaling laws
relating the maximum distance of the chaotic attractor due to digital sampling from the simple periodic
attractor of the continuously sampled system. It is linear
for sufficiently small values of $\tau$ but at larger $\tau$ there is a change in the properties of the
boundedness of the chaotic attractor and the scaling becomes nonlinear.
We will illustrate this effect by considering simple Filippov systems for which the evolution
of a variable $x$ in some region $D\subseteq \RR^n$ is determined by the equations
\begin{equation}
\dot{x}(t) = \begin{cases}
F_1(x(t),\mu) \quad \text{if} \quad H(x(t), \mu) > 0 \\
F_2(x(t),\mu) \quad \text{if} \quad H(x(t), \mu) < 0,
 \end{cases}
\label{sys_eq}
\end{equation}
where $F_1,$ $F_2$ are sufficiently smooth vector functions and $H(x(t),
\mu )$ is some smooth scalar function depending on the system states
$x\in \mathbb{R}^n,$ and parameter $\mu\in\mathbb{R}^m$; $t\in\RR$ is the time variable.
The boundary $\Sigma$ on which $H(x,\mu)=0$ is assumed to be a hyperplane which
divides the region $D$ into two subspaces, $G_1$ (for $H(x,\, \mu) > 0$) and $G_2$ (for $H(x,\, \mu) < 0$), in which the dynamics is
smooth and continuous. There may be extra specifications which determine the motion
across or in $\Sigma$, for example if there is sliding motion as is the case in our
example, but these are standard to include.

In (\ref{sys_eq}) the control of the switching between the two systems across $\Sigma$ is
instantaneous. The modified {\em hybrid} Filippov systems we study are obtained by assuming that
the control function $H$ is evaluated at discrete times $k\tau$, $k=0,1,2,\dots$, for some constant
$\tau >0$, and so the decision to change evolution equation can only occur at these discrete times.
Thus for $k=0,1,2,3,\dots$ we define a discrete variable $i_k$ by
\begin{equation}\label{def_i}
i_{k+1}=\begin{cases}
1 \quad \text{if} \quad H(x(k\tau ), \mu) > 0 \\
2 \quad \text{if} \quad H(x(k\tau ), \mu) < 0 \\
i_k \quad \text{if} \quad H(x(k\tau), \mu) = 0,
\end{cases}
\end{equation}
with $i_{0}=1$ (arbitrarily chosen) so that $i_1$ is always well defined, and replace the evolution
(\ref{sys_eq}) by
\begin{equation}\label{sys_eq1}
\dot x(t)=F_{i_k}(x(t),\mu ) \quad \text{if} \quad (k-1)\tau\le t <k\tau.
\end{equation}

Note that this system excludes the possibility of sliding motion, that is a motion within the discontinuity set $\Sigma$.

\begin{figure}
\centerline{
\begin{picture}(80, 100)
\put(-75, -8){(a)}
\put(-80, 0){\includegraphics[scale = 0.24]{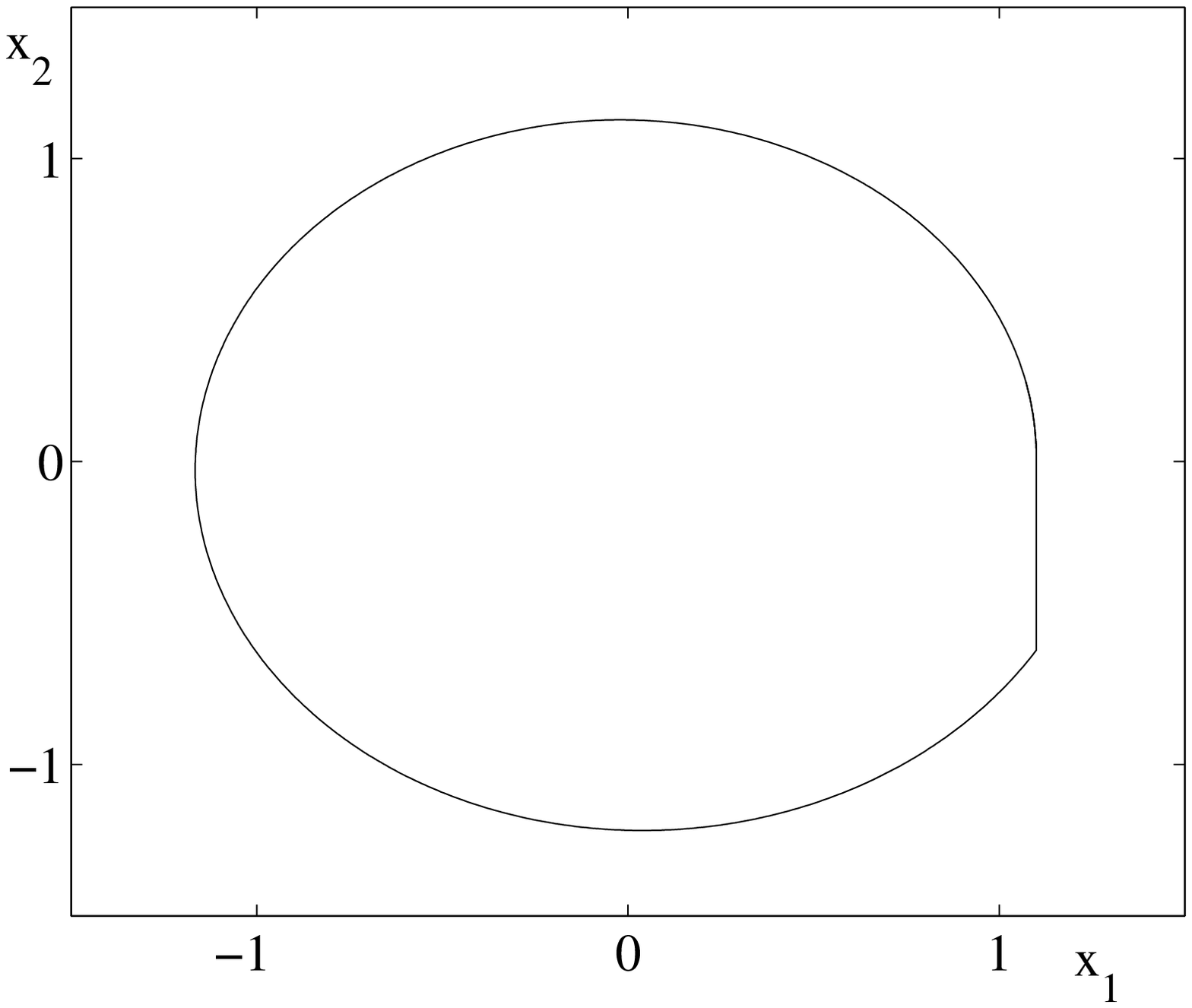}}
\put(45, -8){(b)}
\put(45, 0){\includegraphics[scale = 0.24]{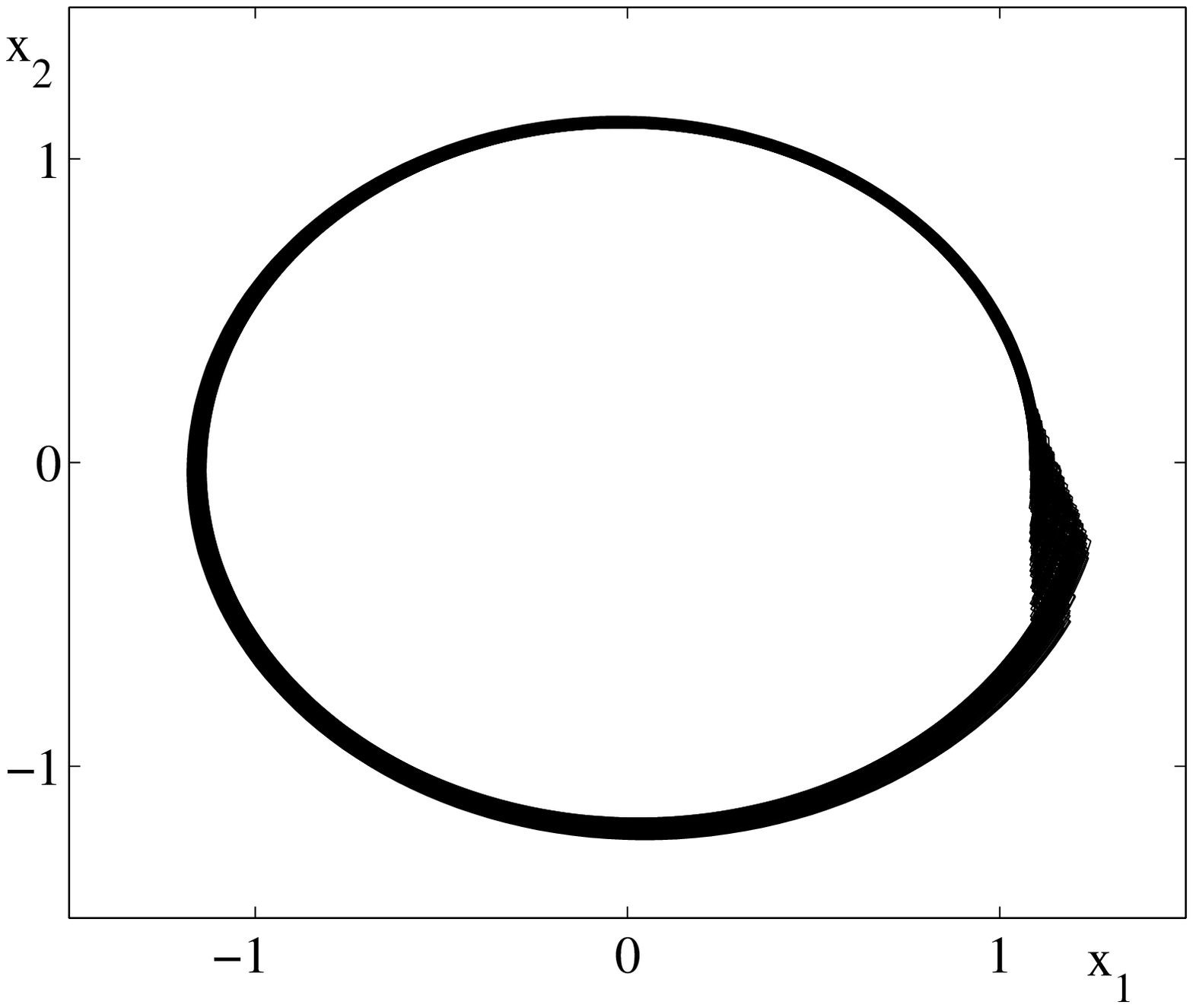}}
\end{picture}
}
\caption{Asymptotic trajectories in Filippov system (\ref{sys_eq1}) without digital sampling (a),
and with digital sampling $\tau = 0.01$ (b).}
\label{fig:1}
\end{figure}

The example which we will consider in the remainder of this paper is planar. Set
\beq\label{sys_eq1}
F_1 = \left\{\begin{array}{c}-\alpha x_1 - \omega x_2 + x_1(x_1^2  + x_2^2)\\
\omega x_1 - \alpha x_2 + x_2 (x_1^2 + x_2^2),
\end{array}\right.
\quad\quad
F_2 = \left\{\begin{array}{c} a \\
b,
\end{array}\right.
\eeq
$H(x) = x_2 - \mu,$ and $h = k\tau$ ($k = 0,\, 1, \, 2,\cdots$)
where $\alpha,$ $\omega$, $a,$ $b,$ $\mu$ and $\tau$ are some chosen constants (system parameters).
The vector field $F_1$ is the normal form for a simple subcritical Hopf bifurcation, with a
stable focus at the origin and an unstable periodic orbit with radius $\sqrt{\alpha}$ if
$\alpha >0$. For appropriate choices of $\mu$, $a$ and $b$ the vector field $F_2$ can be used
to create a stable periodic orbit for the Filippov system (\ref{sys_eq}) as shown in
Figure~\ref{fig:1}(a). The stability is derived from the fact that part of the cycle
lies on $\Sigma$, and this segment of the orbit is called a sliding segment \cite{Fi:88}. This stable cycle
may coexist with the unstable cycle of the vector field $F_1$.

For all the following numerical computations we set $\omega = 15,$ $\alpha = 1.$ In this case the vector
field unstable limit cycle of $F_1$ is centered at the origin and has radius $1.$ The vector field $F_2$
is assumed constant; we set $a = -1$ and $b = 1$. Finally let us set $\mu = 1.1$ so that $\Sigma$
does not intersect the unstable limit cycle of $F_1$. For these parameters (\ref{sys_eq})
with vector fields given by (\ref{sys_eq1}) admits the stable
limit cycle with sliding already referred to as well as the unstable cycle of $F_1$ (the
main restriction in the choice of the parameters is that given $\mu$, the angular velocity $\omega$
in $F_1$ is large enough for the sliding cycle to exist).

Let us now increase the sampling time $\tau$ from $0$ using (\ref{sys_eq1}). As
Fig.~\ref{fig:1}(b) shows, there is an apparent thickening of the
attractor: the stable limit cycle no longer exists and we observe
an onset of more complex asymptotic dynamics. It turns out that this complex dynamics is chaotic and so
there is a transition from a stable orbit to a chaotic attractor due to an introduction of the sampling process.

We will prove this in two parts. First we show that if $\tau >0$ is sufficiently small then there is a
compact set which solutions cannot leave, and hence which contains at least one attractor. This part of the proof is based on showing monotonic crossing of the local transversal by a trajectory, similarly as in the Poincar\'e- Bendixson Theorem \cite{Gl:94}. We present
this argument in fairly general terms below so that the extension to similar systems is clear. Second we show
that any solution in this compact set (and hence the attractor itself) has a positive Lyapunov exponent.

\begin{figure}
\psfrag{a}{(b)}
\psfrag{b}{(a)}
\psfrag{hx}{$\hat{x}$}
\psfrag{xp}{$x^\prime$}
\psfrag{bx}{$\bar{x}$}
\psfrag{mut}{$\mu - \tau$}
\psfrag{uv}{\tiny$(u,\,\,-R)$}
\psfrag{sig}{$\Sigma$}
\psfrag{tx}{$\tilde{x}$}
\psfrag{xg}{$x_g$}
\psfrag{Vr}{$V_R$}
\psfrag{Ur}{$U_R$}
\includegraphics[scale=0.4]{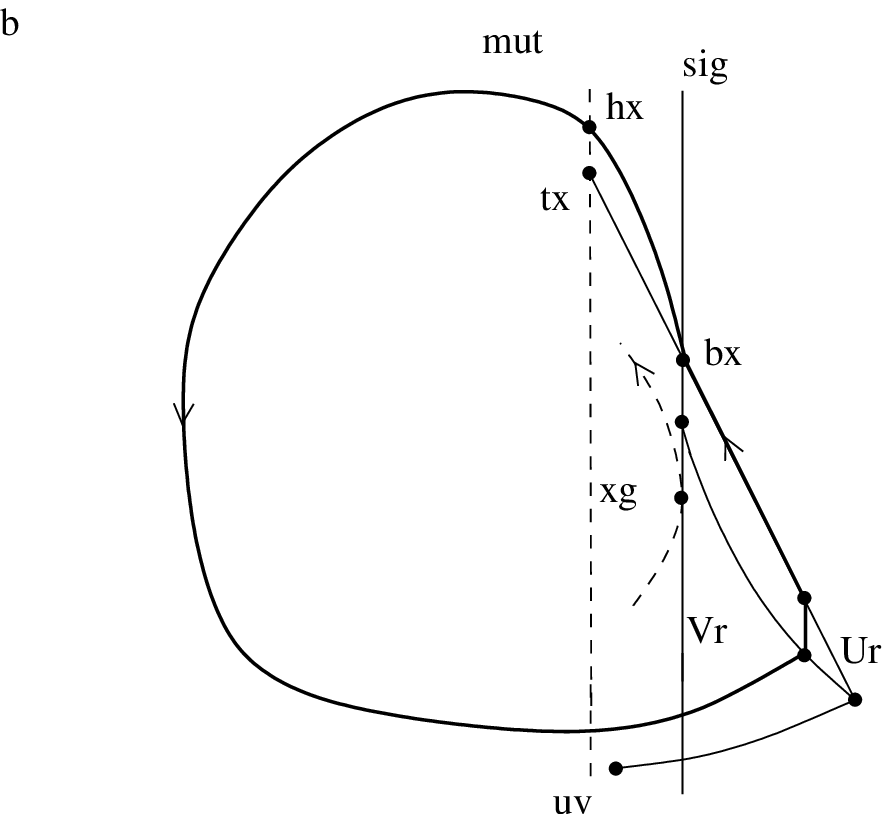}\quad\quad
\includegraphics[scale=0.4]{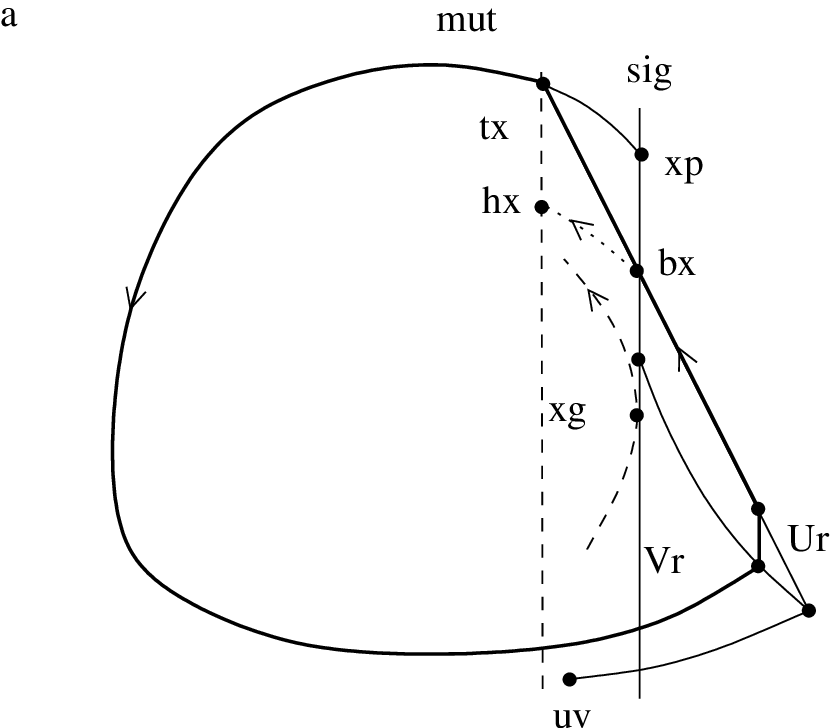}
\caption{Schematic representation of the bounding regions along $\Sigma$ and the boundaries $U_R$ and $V_R$ in the case (a) when $\hat{x} > \tilde{x},$ and (b) $\hat{x} < \tilde{x}.$}
\label{fig:bounD}
\end{figure}

\begin{figure}
\psfrag{tau}{$\tau$}
\psfrag{xb}{$x^b$}
\psfrag{xg}{$x_g$}
\psfrag{xc}{$x^c$}
\psfrag{xp}{$x^p$}
\psfrag{sig}{$\Sigma$}
\psfrag{sigt}{$\Sigma_{F_1}^\tau$}
\psfrag{XB}{$X_B$}
\includegraphics[scale=0.5]{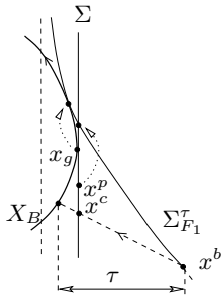}
\caption{Schematic representation of the bounding region $X_B$ for small values of the sampling time $\tau$.}
\label{fig:bounsmalltau}
\end{figure}

The compact invariant region is annular, and its inner boundary is the unstable cycle of $F_1$ with radius one.
To begin the construction of the outer boundary, note that the sliding segment of the (true)
Filippov system terminates at the point $x^g=(\mu ,\, x_2^g)$ which is where the solution of $F_1$ is
tangential to the surface $\Sigma$ as shown in Figures \ref{fig:bounD} and \ref{fig:bounsmalltau} ($x_2^g$ is close to $0$ for $\alpha$ close to $\mu^2$). Now, let $\phi_1(u,v,t)$ denote
the flow generated by $F_1$, i.e. solutions of $F_1$ at time $t$ with initial condition $(u,v)$ and
let $\phi_1^\tau (u,v)=\{(p,q)~|~(p,q)=\phi_1(u,v,t)~\text{for some}~ 0\le t\le \tau\}$. Define
$V$ to be the set of points $(x_1,x_2)$ which can be reached from a point $(u,v)\in G_1\cup\Sigma,$ with
$v\ge -R$ within time $\tau,$
and whose trajectory intersects $G_2$ in time $\tau$, i.e.
the set of points $(x_1,\, x_2)\in G_2$ that are reached from $G_1\cup\Sigma$ within time $\tau$ by following $\phi_1.$
Finally, let $V_R$ be the right hand boundary of $V$, i.e. $(u,v)\in V_R$ such that if $(u^\prime,v)\in V$ then
$u\ge u^\prime$. Along most (and in some examples possibly all) of its length $V_R$ will be the time $\tau$
 image of points on $\Sigma$, but close to $x^g$ this might not be the case. $V_R$ therefore represents
a boundary which no orbit which starts in $G_1$ above $-R$ can cross within time $\tau$ under the flow $\phi_1$.
The choice of $R$ is determined by later considerations, it needs to be large enough to allow the argument
to close up below -- numerical experiments show that $R=2$ is sufficiently large here.

Now consider the effect of the flow generated by $F_2$ to points in $G_2$ to the left of $V_R$. The trajectories
are straight lines with slope $-1$, and as $\dot x_2=-1$ the furthest to the left that an orbit from $G_2$ can reach
in time $\tau$ has $x_1=\mu -\tau$. Let $U$ be the union of $V_R$ and the set of points on straight lines of
slope $-1$ from $(u,v)\in V_R$ with $u>\mu-\tau$ to $\mu -\tau$. Finally let $U_R$
be the right hand boundary of $U$. Note that $U_R$ must be connected.

To summarise: by construction, in time $\tau$, no solution of $F_1$ above the line $x_2=-R$
can move to the right of $V_R$ (which is on the left of $U_R$ or equal to it at places),
and under $F_2$ all such orbits remain to the left of $U_R$ until they return to $G_1$.

Let $\bar{x}$ be the highest point in $U_R$ with $\bar{x}_1=\mu$ and $\tilde{x}$ the highest point in
$U_R$ with $\tilde{x}_1=\mu-\tau$. Let $\hat{x}$ be the first intersection of the solution
of $F_1$ through $\bar{x}$ with $\hat{x}_1 = \mu - \tau$.

If $\hat{x}_2>\tilde{x}_2$ then the outer boundary of the bounding region is the trajectory through $\hat{x}$
under $F_1$ until it hits $V_R$ for the first time (see Figure \ref{fig:bounD}(a)), a horizontal line segment from $V_R$ to
$U_R$, and then $U_R$ back to $\bar{x}$. Note that this requires $R$ to be large enough so that the
trajectory does hit $V_R$. If not then a larger $R$ needs to be chosen.

If $\hat{x}_2<\tilde{x}_2$ let $x^\prime$ be the first preimage of $\tilde{x}$ on $\Sigma$ under $F_1$, and note that
this will lie above $\bar{x}$. Then the outer boundary of the bounding region is the trajectory through $x^\prime$
under $F_1$ until it hits $V_R$ for the first time, a horizontal line segment from $V_R$ to
$U_R$, and then $U_R$ back to $\tilde{x}$ (see Figure \ref{fig:bounD}(b)). As before $R$ needs to be large enough for the connections to work.

In either case we will have created a compact region which no trajectory can exit from, and hence the annular region
contains at least one attractor.


Consider now a sufficiently small $\tau$. Define
$\Sigma_{F_1}^\tau$ as the image of $\Sigma$ under the action of $\phi_1$ for time $\tau.$ Let $x^p\in \Sigma$ be the pre-image of the point at which $\Sigma_{F_1}^\tau$ crosses $\Sigma$ in the neighborhood of $x^g$ as shown in Fig.~ \ref{fig:bounsmalltau}. Define $X_B$ to be the set of initial points $x$ in the neighborhood of $x^g$ such that for any $x\in X_B$ a trajectory generated by $F_1$ evolves through $x^g.$
For $\tau$ sufficiently small $\Sigma_{F_1}^\tau$ is nearly tangent to $\Sigma$ and there exists a point  on $\Sigma_{F_1}^\tau,$ say $x^b,$ such that the
trajectory starting at $x^b$ crosses $\Sigma$ at some point $x^c$ below $x^p,$ and the time of evolution from $x^b\in\Sigma_{F_1}^\tau$ to $X_B$ is $\tau$ (see Fig.~ \ref{fig:bounsmalltau}). Therefore a trajectory starting at $x^b$ must switch to $\phi_1$ to the left of set $X_B$ or on $X_B$. Moreover since $x^p$ lies above $x^c$ no trajectory generated solely by $F_1$ can lie to the right of $X_B$ -- penetrate $G_2$ and return to $G_1$ without switching to $\phi_2.$ We further note that a trajectory rooted at any point within the region bounded by $\Sigma,$ $\Sigma_{F_1}^\tau,$ and the line segment joining $x_b$ with $x_c$ switches to the vector field $F_1$ in a region to the left of $X_B$  -- the time of evolution from any point in this region to reach some point in $G_1$ to the left of $X_B$ is less than $\tau.$ It then follows that for sufficiently small $\tau,$ $X_B$ is a bounding set for the attractor. To find the bounding set to the left of $\Sigma$ we note that the set of points furthest to the left of $\Sigma,$ which can be reached by a trajectory generated by $F_2,$ has co-ordinates $(\mu-\tau, \,\, x_2).$ This implies that along the $x_1$ co-ordinate, in the neighborhood of $x^g,$ the difference between the largest and the smallest values of $x_1$ on the attractor is $\tau.$

For larger values of the sampling time $\tau$ point $x^p$ might no longer lie above $x^c,$ and there exist trajectories in the neighborhood of $x^g$ solely generated by $F_1$ that lie to the right of $X_B.$ In this case the difference between the largest and the smallest values of $x_1$ on the attractor is different from $\tau.$
Therefore, we expect to see a change in the size of the attractor as a function of the sampling time, that changes from a linear law for small $\tau,$ to a different scaling not linearly proportional to $\tau$.

To see that the attractor born out of the stable cycle for non-zero $\tau$ is characterized by a positive
Lyapunov exponent we consider the determinant of the linearized map that maps the neighborhood of $x^g$ onto
itself. Hence, we consider the determinant of the matrix composition of the solutions of the variational
equations for the flows $\phi_1$ and $\phi_2$. Since the vector field $F_2$ that generates the flow $\phi_2$
is a constant vector field then
$$
\det\left(\frac{\partial \phi_2}{\partial x}\right)_t = 1,
$$
where $t$ is the time corresponding to the evolution following $\phi_2,$ and
$\frac{\displaystyle\partial \phi_2}{\displaystyle\partial x} \equiv \Phi_2(t)$ is the fundamental solution matrix corresponding to the
flow $\phi_2$.

Using the explicit expression for the flow function $\phi_1$ in the polar co-ordinates we can compute the fundamental solution matrix $\Phi_1,$ corresponding to the flow $\phi_1:$
$$
\Phi_1(t_j) \equiv \left(\frac{\partial \phi_1}{\partial (\rho,\,\,\theta)}\right)_{t_j} = \left(\begin{array}{cc}  f_j(t_j,\rho_j)\exp(2t_j) & 0 \\ 0 & 1
\end{array}\right),
$$
 with $j = 0,\,1,\,2\cdots$ that correspond to the times of evolution following $\phi_1$ after
 $j-$th switching from the flow $\phi_2$ to $\phi_1.$ At $j = 0$ we initialize the evolution from
 the neighborhood of $x^g,$ and $\rho_j$ denotes the radius from the origin at the $j-$th switching instance. Finally $f_j(0,\rho_j) = 1$ and $f_j$ are monotonically increasing functions of $t_j$.

Therefore,
$$
\det \Phi_1(t_j) = \det\left(\frac{\partial \phi_1}{\partial (\rho,\,\theta)}\right)_{t_j} > 1,
$$
and $\sum_j t_j > \frac{\pi}{\omega}$ (note that the flow follows $\phi_1$ from $x^g$ until the first intersection with $\Sigma$ for the amount of time greater than $\pi/\omega$).
The resulting determinant of the composition of the fundamental solution matrices corresponding to
the flows $\phi_1$ and $\phi_2$ is
$\Pi_{j=0}^{n-1}\det \Phi_1(t_j)\times 1$
where $n$ is the number of switchings that are required for the system trajectory to reach the
neighborhood of $x^g.$ We should note here that formally we should introduce a co-ordinate transformation to the flow $\phi_2$ and use the fundamental solution matrix corresponding to the vector field $F_2$ in polar co-ordinates. However, since the flow $\phi_2$ is a constant flow the determinant of $\Phi_2$ is always $1$ regardless on the co-ordinate set.

Therefore, we are only interested in $\Pi_{j=0}^{n-1}\det \Phi_1(t_j).$
Since this product is greater than $1$ and the determinant of a matrix is
the product of its eigenvalues we conclude that there is at least one eigenvalue of $\Pi_{j=0}^{n-1}\Phi_1(t_j)$
which is characterized by the magnitude greater than one. This eigenvalue is the exponential of the Lyapunov exponent.

 Let us determine the size of the attractor, measured as the distance
from the largest to smallest values which the attractor attains on a Poincar\'e section defined on
a set $\{x_2 = 0, - 1.3 < x_1 < -1\}.$
For sufficiently small values of the sampling time $\tau$ the size of the attractor along the $x_1$ co-ordinate is proportional to $\tau$ around the point $x^g.$ However, the size of
the attractor is measured on section $\{x_2 = 0, - 1.3 < x_1 < -1\}.$ Therefore, we have to determine the expansion of the attractor along the flow after time $\pi/\omega$ (which is
the time required to map the points from the neighborhood of $x^g$ using flow $\phi_1$ onto our chosen
Poincar\'e section). This expansion is captured by the non-trivial Floquet multiplier of the fundamental solution matrix. To find this multiplier we use the explicit solutions of
the differential equations that define $F_1$ in polar co-ordinates. We find that
$$
\rho(t) = \sqrt{\frac{\alpha}{(1-\alpha\rho_0^{-2})\exp(2\alpha t) - 1}},\quad\theta = \theta_0 + \omega t.
$$
Differentiating $\rho(t)$ with respect to $\rho_0,$ and after substituting for $\alpha = 1,$  $t = \pi/15,$ and for $\rho_0 = 1.1$ we get
 $
\frac{\displaystyle d\rho}{\displaystyle d \rho_0} = 1.8084.
 $
 \begin{figure}
\centerline{\includegraphics[scale=0.4]{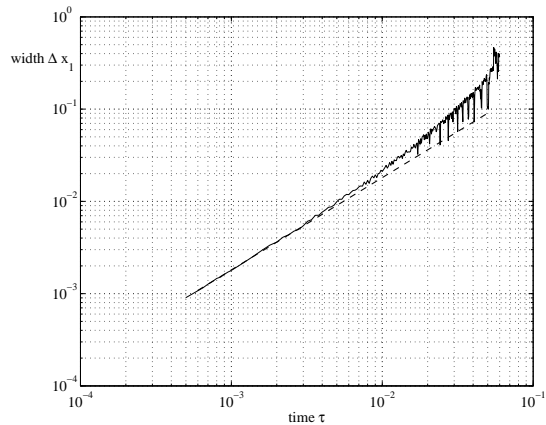}
}
\caption{Size of the attractor versus the sampling time $\tau.$ The dashed line refers to theoretical predictions of the size of the attractor.}
\label{fig:scaling}
\end{figure}
In Fig.~\ref{fig:scaling} using the logarithmic scales we are depicting how the size of the attractor scales against the sampling time $\tau.$ The dashed diagonal line
refers to the linear scaling proportional to $\tau$ obtained using the above theoretical prediction. We can see that the theoretical prediction coincides with the numerical results for small values of $\tau.$ For $\tau$ sufficiently small the attractor is bounded by the trajectory leaving $\Sigma$ at $x^g.$

We notice that the increase in the value of the sampling time $\tau$ above $\tau = 0.01$ results in the growing discrepancy between the numerical and theoretical values. This comes from the fact that the attractor is no longer bounded by the trajectory leaving $\Sigma$ at $x^g.$ Other effects such as resonances between the sampling time $\tau$ and the rotation $\omega$ of the flow $\phi_1$ produce other local variations in the scaling law visible in Fig.~\ref{fig:scaling}.

In conclusion, we have shown on a planar example that the digital sampling applied to the decision function in Filippov type systems leads to the onset of chaotic dynamics. Further on we have shown that for sufficiently small values of the sampling time $\tau$ the size of the chaotic attractor scales linearly with the sampling time $\tau.$ The mechanism that leads to the onset of chaos is triggered by the expansion of the volume of phase space produced by the flow $\phi_1$ combined with the
re-injection triggered by the subsequent application of the switchings along the manifold $\Sigma.$ The scenario observed in our model example will be present in a larger class of systems with swithings. The essential ingredients of these systems will be the presence of a stable cycle (when no digitization is applied) and the presence of an expansion of a volume of phase space in the presence of digitization.

The onset of chaotic dynamics triggered by this mechanism is similar to an abrupt transition from a stable periodic orbit with sliding to a small scale chaotic dynamics that might occur in Filippov type systems under an introduction of an arbitrarily small time delay in the switching function \cite{Si:06}. On the practical side, in spite of the fact that these oscillations are micro-chaotic, they can can be highly harmful to control elements, induce excessive wear to machine tools \cite{EnSt:98}, and therefore it is important to understand the mechanisms that might trigger this type of complex dynamics.

Research partially funded by EPSRC grant EP/E050441/1 and the University
of Manchester.


\end{document}